\documentclass[aps,prd,twocolumn,superscriptaddress,amsmath,amssymb,nofootinbib,longbibliography,preprintnumbers,floatfix]{revtex4-2}
\usepackage[dvipsnames]{xcolor}
\usepackage[colorlinks]{hyperref}
\usepackage[normalem]{ulem}  
\usepackage{xspace}

\usepackage{todonotes}

\usepackage{tikz}
\usetikzlibrary{tikzmark}
\usetikzlibrary{positioning}
\usepackage[normalem]{ulem} 

\hypersetup{linkcolor=BrickRed,citecolor=Green,
filecolor=Mulberry,
urlcolor=NavyBlue,
menucolor=BrickRed,
runcolor=Mulberry
}

\definecolor{lime}{HTML}{A6CE39}
\DeclareRobustCommand{\orcidicon}{\hspace{-1mm}
	\begin{tikzpicture}
	\draw[lime, fill=lime] (0,0) 
	circle [radius=0.16] 
	node[white] {{\fontfamily{qag}\selectfont \tiny \,ID}};
	\draw[white, fill=white] (-0.0525,0.095) 
	circle [radius=0.007];
	\end{tikzpicture}
	\hspace{-3mm}
}

\foreach \x in {A, ..., Z}{\expandafter\xdef\csname orcid\x\endcsname{\noexpand\href{https://orcid.org/\csname orcidauthor\x\endcsname}
			{\noexpand\orcidicon}}
}

\usepackage{graphicx}
\usepackage{bbm}

\makeatletter
\newcommand{\abs}{\@ifstar\abssmall\absbig}
\newcommand{\absbig}[1]{\left \lvert #1 \right \rvert}
\newcommand{\abssmall}[1]{\lvert #1 \rvert}
\makeatother

\renewcommand{\i}{\mathrm{i}}
\newcommand{\Tr}{\mathrm{Tr}}
\renewcommand{\Im}{\mathrm{Im}}
\renewcommand{\Re}{\mathrm{Re}}
\newcommand{\dd}{\mathrm{d}}

\newcommand{\vrho}{\varrho}
\newcommand{\bvrho}{\bar{\varrho}}
\newcommand{\bN}{\overline{N}}
\newcommand{\bF}{\overline{F}}
\newcommand{\veps}{\varepsilon}
\newcommand{\Id}{\mathbb{I}}

\begin{document}
\preprint{N3AS-24-031}

\title{Quantum maximum entropy closure for small flavor coherence}

\author{Julien Froustey\orcidA{}}
\email{jfroustey@berkeley.edu}
\affiliation{Department of Physics, University of California Berkeley, Berkeley, CA 94720, USA}
\affiliation{Department of Physics, North Carolina State University, Raleigh, NC 27695, USA}
\affiliation{Department of Physics, University of California San Diego, La Jolla, CA 92093, USA}

\author{James P. Kneller\orcidB{}}
\affiliation{Department of Physics, North Carolina State University, Raleigh, NC 27695, USA}

\author{Gail C. McLaughlin\orcidC{}}
\affiliation{Department of Physics, North Carolina State University, Raleigh, NC 27695, USA}

\begin{abstract}
    Quantum angular moment transport schemes are an important avenue toward describing neutrino flavor mixing phenomena in dense astrophysical environments such as supernovae and merging neutron stars. Successful implementation will require new closure relations that go beyond those used in classical transport. In this paper, we derive the first analytic expression for a quantum M1 closure, valid in the limit of small flavor coherence, based on the maximum entropy principle. We verify that the resulting closure relation has the appropriate limits and characteristic speeds in the diffusive and free-streaming regimes. We then use this new closure in a moment linear stability analysis to search for fast flavor instabilities in a binary neutron star merger simulation and find better results as compared with previously designed, \emph{ad hoc}, semiclassical closures.
\end{abstract}

\maketitle

\section{Introduction}

A key ingredient in simulations of core-collapse supernovae (CCSNe) or neutron star mergers (NSMs) is the neutrino transport~\cite{Mezzacappa:2020oyq,Foucart:2022bth,Fischer:2023ebq,Wang:2023tso}, with a large body of literature dedicated to the description of radiation transport through a reduced set of angular moments~\cite{Thorne:1981nvt,Shibata:2011kx,Cardall:2012at}. The so-called M1 scheme, which evolves the number density and number flux instead of the full angular distribution of neutrinos, requires an appropriate \emph{closure} to obtain a closed system of equations~\cite{Levermore_1984,Smit_closure_2000,Murchikova:2017zsy,Richers:2020ntq}. Following the seminal work of Minerbo~\cite{Minerbo_1978}, generalized for fermionic radiation in~\cite{Cernohorsky_1989,Cernohorsky_closure_1994}, analytic closures based on the maximum entropy principle have been developed for “classical” radiation, i.e., without taking into account the phenomenon of flavor mixing.

However, neutrino flavor transformation is expected to occur in CCSN and NSM environments with important consequences on the dynamics. Flavor transformation can occur due to various mechanisms including MSW transitions, turbulence, matter-neutrino resonances, as well as the so-called “slow,” “fast,” and collisional instabilities (see, e.g.,~\cite{Wolfenstein:1977ue,Mikheyev:1985zog,Loreti:1995ae,Kneller:2010sc,Duan:2010bg,Malkus:2012ts,Wu:2015fga,Horiuchi:2018ofe,Tamborra:2020cul,Johns:2021qby,Capozzi:2022slf,Richers_review,Volpe:2023met,Sen:2024fxa} and many references therein). In order to include these mechanisms in large-scale simulations, beyond the studies using hybrid~\cite{2020PhRvD.102h1301S} or phenomenological approaches to describe flavor transformation~\cite{Li:2021vqj,Just:2022flt,Ehring:2023lcd,Ehring:2023abs}, moment schemes must be extended to account for flavor mixing. As in the classical case, an appropriate closure is needed, since a simple truncation of the tower of quantum moments cannot accurately describe flavor evolution~\cite{Johns:2019izj,Johns:2020qsk}. In recent studies~\cite{Myers:2021hnp,Grohs:2022fyq,Grohs:2023pgq,Froustey:2023skf} it has been shown that dynamical calculations of flavor transformation with quantum moments are able to capture the qualitative and even quantitative features found in multiangle calculations (e.g.,~\cite{Padilla-Gay:2020uxa,Richers:2022bkd,Nagakura:2022qko,Nagakura:2023xhc,Xiong:2024tac,Richers:2024zit}) with discrepancies that have been attributed to the \emph{ad hoc}, semiclassical closures that were used.

In this paper, we derive the first \emph{ab initio} quantum closure for angular moments by applying the maximum entropy principle to the one-body reduced density matrix $\vrho$, which generalizes the classical neutrino distribution functions. We restrict ourselves to the case of small flavor coherence, so that the flavor diagonal elements of $\vrho$ follow (to leading order) the classical maximum entropy closure (MEC), and we obtain the correction describing the closure for the off-diagonal elements of $\vrho$. For clarity, we present in Sec.~\ref{sec:deriv_MEC} our derivation for particles that follow Maxwell-Boltzmann statistics (see also details in Appendix~\ref{app:extremization}), and derive in Appendix~\ref{app:FermiDirac} the result for particles with Fermi-Dirac statistics. Our presentation is also limited to the case of axially symmetric systems in momentum space, so that there is only one nonzero component of the flux vector which is along the symmetry axis (that we call $z$) $F^{z}$, and so that the symmetric components of the pressure tensor are $P^{zz}$ and $P^{xx}=P^{yy}=(N-P^{zz})/2$, with $N$ the number density moment. Mathematical and physical requirements on the closure are discussed in Sec.~\ref{sec:requirements} and Appendices~\ref{app:free_streaming} and~\ref{app:charac_speeds}. In Sec.~\ref{sec:LSA}, we illustrate the promise of this new quantum closure by incorporating it into the moment linear stability analysis framework of~\cite{Froustey:2023skf}, focusing on fast flavor instabilities in a classical NSM simulation and show how it improves the agreement of the predictions with multiangle linear stability analysis.

\section{Derivation of the Maximum Entropy Closure}
\label{sec:deriv_MEC}

\subsection{Classical MEC}

Before generalizing to a quantum system, we first outline the derivation in the classical case.
The Maxwell-Boltzmann maximum entropy closure, or “Minerbo” closure~\cite{Minerbo_1978}, is obtained by maximizing the angular entropy under the constraints of given number density and flux. Writing as $g(\mu)$ the angular distribution for an axisymmetric system, with $\mu$ the cosine of the polar angle, the following functional is maximized:
\begin{multline}
\label{eq:functional_class}
    \mathsf{S}[g] = \int_{-1}^{1}{\dd \mu \, g \, \ln(g)} - \alpha \left(\int_{-1}^{1}{\dd \mu \, g} - \frac{N}{2 \pi} \right) \\ - Z \left(\int_{-1}^{1}{\dd \mu \,\mu \,g} - \frac{F^z}{2 \pi}\right) \, ,
\end{multline}
where $\alpha$ and $Z$ are two Lagrange multipliers. Maximizing with respect to $g(\mu)$ leads to the functional form $g(\mu) = e^{\alpha -1} e^{Z \mu}$, which, after inserting it in the two constraints (number density and flux), reads
\begin{equation}
\label{eq:distrib_ME}
    g(\mu) = \frac{N}{4 \pi} \frac{Z}{\sinh(Z)} e^{Z \mu} \, ,
\end{equation}
with
\begin{equation}
\label{eq:Z_ME}
    \coth(Z) - \frac{1}{Z} = \frac{F^z}{N} \equiv f \, .
\end{equation}
M1 transport requires providing the relation of the pressure tensor\footnote{Technically, the second moment of the number distribution does not have the units of pressure (which is normally defined as the second moment of the \emph{energy} density), but we stick to the name for simplicity.} $P^{zz} = 2 \pi \int_{-1}^{1}{\dd \mu \, \mu^2 \, g(\mu)}$ to the number density and flux, which reads for the Minerbo closure $P^{zz} = N - 2F^{z}/Z \equiv \chi(f) N$. Note that the nonnegative nature of the distribution function leads to the properties $N \geq 0$ and $\lvert F^z \rvert / N \leq 1$, which are conversely the conditions for the maximum entropy optimization problem~\eqref{eq:functional_class} to have a solution~\cite{LareckiBanach}.

\subsection{Quantum MEC}
We now seek to generalize Eq.~\eqref{eq:functional_class}, for a density matrix $\vrho$ where the entries of the matrix are the occupation numbers (diagonal elements, generalizing the previous distributions $g$) and coherences (off-diagonal elements) of the flavor quantum states~\cite{Sigl:1993ctk,Volpe:2013uxl,Vlasenko:2013fja}.
Using the von Neumann expression for the entropy~\cite{vonNeumann_1927,von2018mathematical}, we write the entropy functional with constraints as
\begin{multline}
\label{eq:functional_quant}
    \mathsf{S}[\vrho] = \int_{-1}^{1}{\dd \mu \, \Tr[\vrho \ln(\vrho)]} - \left \langle \alpha, \int_{-1}^{1}{\dd \mu \, \vrho} - \frac{N}{2 \pi} \right\rangle \\ - \left \langle Z, \int_{-1}^{1}{\dd \mu \, \mu \, \vrho} - \frac{F^z}{2 \pi}\right\rangle \, ,
\end{multline}
where the angle brackets indicate the Frobenius inner product between two Hermitian matrices $A$ and $B$, $\langle A,B \rangle = \Tr(A B^\dagger)$, and $\alpha$ and $Z$ are two Hermitian matrices of Lagrange multipliers.

We now assume that coherences in $\vrho$ are small i.e., $\abs{\vrho_{a \neq b}} \ll \vrho_{aa}$. For brevity, we shall consider a two-flavor system so that 
\begin{equation}
    \vrho = \begin{pmatrix} \vrho_{ee} & \vrho_{ex} \\
    \vrho_{ex}^* & \vrho_{xx} \end{pmatrix} \quad \text{with} \quad \abs{\vrho_{ex}} \ll \vrho_{ee}, \vrho_{xx} \, ,
\end{equation}
but our results are easily generalized to three or more flavors. 
For the $\ln(\vrho)$ that appears in Eq.~\eqref{eq:functional_quant} we use the expansion~\cite{MatrixLog_Adler,MatrixLog_Haber}
\begin{multline}
    \ln(A+tB) = \ln(A) + t \int_{0}^{\infty}{\dd s \frac{1}{A + s \Id}B\frac{1}{A + s \Id}} \\
    - t^2 \int_{0}^{\infty}{\dd s \frac{1}{A + s \Id}B\frac{1}{A+s \Id}B\frac{1}{A+s \Id}} + \mathcal{O}(t^3) \, ,
\end{multline}
with $\Id$ the identity matrix and $t$ an expansion parameter that represents the relative size of the off- and on- diagonal components of $\vrho$. When we set $A = \mathrm{diag}(\vrho_{ee},\vrho_{xx})$ and $tB = \left(\begin{smallmatrix} 0 & \vrho_{ex} \\ \vrho_{ex}^* & 0 \end{smallmatrix}\right)$ we find, at zeroth order,
\begin{equation}
\label{eq:log_rho_0}
    \ln(\vrho)^{(0)} = \begin{pmatrix} \ln(\vrho_{ee}) & 0 \\
    0 & \ln(\vrho_{xx}) \end{pmatrix} \, .
\end{equation}
At first order, we get
\begin{align}
    \ln(\vrho)^{(1)} &= \int_{0}^{\infty}{\dd s \begin{pmatrix} 0 & \frac{\vrho_{ex}}{(\vrho_{ee}+s)(\vrho_{xx}+s)} \\
    \frac{\vrho_{ex}^*}{(\vrho_{ee}+s)(\vrho_{xx}+s)} \end{pmatrix}
    } \nonumber \\
    &= \frac{\ln(\vrho_{ee}) - \ln(\vrho_{xx})}{\vrho_{ee} - \vrho_{xx}} \begin{pmatrix} 0 & \vrho_{ex} \\ \vrho_{ex}^* & 0 \end{pmatrix} \, ,
\end{align}
and, finally, at second order, we find
\begin{align}
    \ln(\vrho)^{(2)} &= - \int_{0}^{\infty}{\dd s \begin{pmatrix} \frac{\abs{\vrho_{ex}}^2}{(\vrho_{ee}+s)^2(\vrho_{xx}+s)} & 0 \\
    0 & \frac{\abs{\vrho_{ex}}^2}{(\vrho_{ee}+s)(\vrho_{xx}+s)^2} \end{pmatrix}
    } \nonumber \\
    &= - \abs{\vrho_{ex}}^2 \begin{pmatrix} D_{e,x} & 0 \\ 0 & D_{x,e} \end{pmatrix} \, ,
\end{align}
with
\begin{equation}
    D_{a,b} \equiv \frac{\vrho_{aa}\left[\ln(\vrho_{aa}) - \ln(\vrho_{bb})\right] - \left(\vrho_{aa}-\vrho_{bb}\right)}{\vrho_{aa} (\vrho_{aa} - \vrho_{bb})^2} \, .
\end{equation}
Combining all these contributions, we get
\begin{multline}
\label{eq:Tr_rhologrho}
    \Tr[\vrho \ln(\vrho)] = \vrho_{ee} \ln(\vrho_{ee}) + \vrho_{xx} \ln(\vrho_{xx}) \\ + \frac{\ln(\vrho_{ee})-\ln(\vrho_{xx})}{\vrho_{ee} - \vrho_{xx}} \abs{\vrho_{ex}}^2 + \cdots
\end{multline}
For three or more flavors this result generalizes to
\begin{multline}
    \Tr[\vrho \ln(\vrho)] = \sum_a \vrho_{aa} \ln(\vrho_{aa}) \\ + \frac12 \sum_a \sum_{b\neq a} \frac{\ln(\vrho_{aa})-\ln(\vrho_{bb})}{\vrho_{aa} - \vrho_{bb}} \abs{\vrho_{ab}}^2 + \cdots 
\end{multline}
The factor $1/2$ accounts for the two identical terms in the sum, as $\vrho_{ba} =\vrho_{ab}^*$.

At leading order, maximizing \eqref{eq:functional_quant} with respect to $\vrho_{aa}$ gives the same result as Eqs.~\eqref{eq:distrib_ME}--\eqref{eq:Z_ME}. We write as $Z_{aa}$ the solution of Eq.~\eqref{eq:Z_ME} with $f_{aa}=F_{aa}^z/N_{aa}$. The diagonal elements of the pressure tensor then read $P_{aa}^{zz} = N_{aa} - 2\,F^{z}_{aa} / Z_{aa} = \chi(f_{aa}) N_{aa}$. For $\vrho_{ex}$ on the other hand, we get\footnote{Rigorously, we maximize over $\Re(\vrho_{ex})$ and $\Im(\vrho_{ex})$ separately, the equations being then combined to give~\eqref{eq:distrib_QMbo}—see details in Appendix~\ref{app:extremization}.}
\begin{equation}
\label{eq:distrib_QMbo}
    \vrho_{ex}(\mu) = \frac{\vrho_{ee}-\vrho_{xx}}{\ln(\vrho_{ee})-\ln(\vrho_{xx})} \left(\alpha_{ex} + \mu Z_{ex}\right) \, ,
\end{equation}
with two complex numbers $\alpha_{ex}$ and $Z_{ex}$ such that $2 \pi \int_{-1}^{1}{\dd \mu \vrho_{ex}} = N_{ex}$ and $2 \pi \int_{-1}^{1}{\dd \mu \mu \vrho_{ex}} = F_{ex}^z$. Note that the expression~\eqref{eq:distrib_QMbo} is \emph{not} singular for $\vrho_{ee}(\mu) = \vrho_{xx}(\mu)$. In that situation, since $\lim_{s \to 1} (1-s)/\ln(s) = 1$, we have $\vrho_{ex}(\mu) = \vrho_{ee}(\mu) \left(\alpha_{ex} + \mu Z_{ex}\right)$. For three or more flavors, the same formula is valid for any off-diagonal element $\vrho_{ab}$, with $(\vrho_{ee},\vrho_{xx}) \to (\vrho_{aa},\vrho_{bb})$ in Eq.~\eqref{eq:distrib_QMbo}.

Since $\vrho$ is, by definition, a positive semidefinite matrix, its diagonal components are non-negative numbers and the classical results for maximum entropy distributions (e.g.,~\cite{LareckiBanach}) directly apply. Even for large flavor coherence, we have $N_{aa} \geq 0$ and $\lvert f_{aa} \rvert \leq 1$. The parameter $Z_{aa}$ can be obtained by any method inverting the Langevin function~\eqref{eq:Z_ME} and having the appropriate limits, see, e.g.,~\cite{Cernohorsky_closure_1994,jedynak2017new} for approximate formulas.

We now introduce the integrals $I_{(n)}$, which depend on the classical maximum entropy distributions~\eqref{eq:distrib_ME} for $\vrho_{ee}(\mu)$ and $\vrho_{xx}(\mu)$:
\begin{equation}
\label{eq:In}
    I_{(n)} \equiv 2 \pi \int_{-1}^{1}{\dd \mu \, \mu^n \frac{\vrho_{ee} - \vrho_{xx}}{\ln(\vrho_{ee}) - \ln(\vrho_{xx})}} \, .
\end{equation}
These quantities allow us to succinctly express the off-diagonal elements of the moments as $N_{ex} = I_{(0)} \, \alpha_{ex} + I_{(1)} \, Z_{ex}$, $F_{ex}^z = I_{(1)} \, \alpha_{ex} + I_{(2)} \, Z_{ex}$ and $P_{ex}^{zz} = I_{(2)} \,\alpha_{ex} + I_{(3)} \, Z_{ex}$. The equations for the first two moments can be inverted to get
\begin{equation}
\label{eq:alpha_Z}
\begin{aligned}
    \alpha_{ex} &= \frac{I_{(2)} N_{ex} - I_{(1)} F_{ex}^z}{I_{(0)} I_{(2)} - I_{(1)}^2} \, , \\
    Z_{ex} &= \frac{I_{(0)} F_{ex}^z - I_{(1)} N_{ex}}{I_{(0)}I_{(2)} - I_{(1)}^2} \, ,
\end{aligned}
\end{equation}
allowing us to write the equation for the off-diagonal element of the $zz$ entry of the pressure tensor as
\begin{equation}
\label{eq:closure_QMbo}
    P_{ex}^{zz} = \kappa N_{ex} + \eta F_{ex}^z \, ,
\end{equation}
where
\begin{equation}
\label{eq:kappa_eta}
\begin{aligned}
    \kappa &\equiv \frac{I_{(2)}^2 - I_{(1)}I_{(3)}}{I_{(0)}I_{(2)} - I_{(1)}^2} \, , \\
    \eta &\equiv \frac{I_{(0)}I_{(3)} - I_{(1)}I_{(2)}}{I_{(0)}I_{(2)} - I_{(1)}^2} \, .
\end{aligned}
\end{equation}
Note that $h(\mu) \equiv [\vrho_{ee}-\vrho_{xx}]/[\ln(\vrho_{ee})-\ln(\vrho_{xx})]$ is a positive function. This allows us to see $I_{(n)}$ as the moments of a random variable, with probability density $2\pi\, h(\mu)/I_{(0)}$. The positivity of the variance means that $I_{(0)}I_{(2)} \geq I_{(1)}^2$. This inequality is only an equality in the limit of flux factors exactly equal to $1$, a case we treat explicitly below. Equation~\eqref{eq:alpha_Z} can thus be generally solved for any off-diagonal moments $(N_{ex},F_{ex}^z)$ in the small flavor coherence regime. For three of more flavors, the previous equations are true for each off-diagonal component $(a,b)$, defining the $I_{(n)}$ integrals with $\vrho_{aa}$ and $\vrho_{bb}$ in place of $\vrho_{ee}$ and $\vrho_{xx}$.

Equations~\eqref{eq:distrib_QMbo} and~\eqref{eq:closure_QMbo} are the main results of this paper. Along with Eqs.~\eqref{eq:distrib_ME}--\eqref{eq:Z_ME} for the diagonal elements of $\vrho$, all the elements of the pressure moment are now defined and thus we have obtained a quantum generalization of the Minerbo closure. Using the maximum entropy principle for small flavor coherence, we showed that the flavor off-diagonal pressure tensor is a linear combination of the associated number density and spatial flux density, with real coefficients determined by the classical distributions for $\nu_e$ and $\nu_x$.

\subsection{Maximum-entropy Fermi-Dirac closure}
The very same strategy can be followed taking into account the fermionic nature of neutrinos~\cite{Cernohorsky_1989,Cernohorsky_closure_1994}. The starting ansatz is that the entropy part of the functional \eqref{eq:functional_quant} becomes $\int_{-1}^{1}{\dd \mu \, \Tr[\vrho \ln(\vrho) + (\mathbb{I}-\vrho)\ln(\mathbb{I}-\vrho)]}$, see, e.g.,~\cite{Sigl:1993ctk}. The mathematical details are more complex and can be found in Appendix~\ref{app:FermiDirac}. The end result is that Eq.~\eqref{eq:closure_QMbo} keeps the same form, although the expressions for the integrals $I_{(n)}$ given in Eq.~\eqref{eq:In} are modified such that $\ln(\vrho_{aa}) \to \ln( \vrho_{aa}/(1-\vrho_{aa}))$.

\section{Physical requirements on the closure}
\label{sec:requirements}

For classical neutrino transport, the non-negativity of the distribution function leads to various constraints that must be satisfied by any physical closure, notably $f^2 \leq \chi \leq 1$, with the limits $\lim_{f \to 0} \chi = 1/3$ (diffusive regime) and $\lim_{f \to 1} \chi = 1$ (free-streaming regime). These constraints are satisfied by the classical Minerbo closure~\cite{Anile_EddingtonFactors_1991,Smit_closure_2000} and therefore by the diagonal elements of our quantum Minerbo closure. 
We now show that the off-diagonal elements of the quantum Minerbo closure also satisfy such constraints in the same, appropriately defined, limits. 
Note however that, since $\vrho_{ex}$ is a complex number, $N_{ex}$ is not necessarily positive, and $\abs{F_{ex}^z/N_{ex}}$ does not have to be smaller than $1$.

If the flavor-diagonal distributions are isotropic, the integrals~\eqref{eq:In} can be calculated explicitly, with $I_{(2k)} = 1 / (2k+1) \times [N_{ee}-N_{xx}]/[\ln(N_{ee})-\ln(N_{xx})]$ and $I_{(2k+1)} = 0$ for $k \in \mathbb{N}$. As a consequence, we find $\kappa = 1/3$ and $\eta = 0$ and so  Eq.~\eqref{eq:closure_QMbo} reads $P_{ex}^{zz} = N_{ex}/3$, regardless of the value of $F_{ex}^{z}$. This is not a trivial result, since it is possible to have $\vec{F}_{aa} = \vec{0}$ but $\vec{F}_{ex} \neq \vec{0}$, an example being the “isotropy-breaking modes” of collisional instabilities~\cite{Liu:2023pjw}. 

On the opposite end, in the free-streaming regime, let us assume that\footnote{We cannot deal with the case where $\nu_e$ and $\nu_x$ would free-stream in different directions, since, in the direction of $\nu_e$, the $xx$ entry of $\vrho$ would be zero, making the logarithm~\eqref{eq:log_rho_0} non-defined and the whole expansion procedure breaks down.} $F_{ee}^z/N_{ee} = F_{xx}^z/N_{xx}=1$. Then $I_{(n)} = [N_{ee}-N_{xx}]/[\ln(N_{ee})-\ln(N_{xx})] \equiv I \, \forall \, n \in \mathbb{N}$. Equations~\eqref{eq:alpha_Z} and~\eqref{eq:kappa_eta} appear singular, but we can show (see Appendix~\ref{app:free_streaming}) that in this limit $\kappa \to -1$ and $\eta \to 2$. In addition, the constraint equations only have a solution if $F_{ex}^z = N_{ex}^z = I(\alpha_{ex} + Z_{ex})$, in which case we also have $P_{ex}^{zz} = N_{ex}$. 

Another physical constraint on the closure emerges from causality. 
Just as Boltzmann equations can be rewritten in a classical M1 two-moment scheme, the Quantum Kinetic Equations\footnote{Note that this approach neglects potential many-body effects that could have consequences on neutrino evolution in dense environments, see e.g.,~\cite{Patwardhan:2022mxg}.} (QKEs) describing the evolution of $\vrho$ can be transposed in a “quantum” two-moment scheme~\cite{Richers:2019grc,Myers:2021hnp,Grohs:2022fyq,Grohs:2023pgq,Froustey:2023skf,Kneller:2024buy}
\begin{equation}
    \begin{aligned}
        \frac{\partial N}{\partial t} + \frac{\partial F^z}{\partial z} &= \mathcal{S}_N \, , \\
        \frac{\partial F^z}{\partial t} + \frac{\partial P^{zz}}{\partial z} &= \mathcal{S}_F \, ,
    \end{aligned}
\end{equation}
where $\mathcal{S}_{N,F}$ are the source terms, i.e., moments of the Hamiltonian-like and collision terms appearing in the multiangle QKE. We write this system of (flavor) matrix equations in the compact form~\cite{Audit:2002hr,Pons:2000br}
\begin{equation}
\label{eq:QKE}
    \frac{\partial \mathcal{U}}{\partial t} + \frac{\partial \mathcal{F}}{\partial z} = \mathcal{S}(\mathcal{U}) \, ,
\end{equation}
where
\begin{equation}
\mathcal{U} = \begin{pmatrix}
N_{ee} \\ F_{ee}^{z} \\ N_{xx} \\ F_{xx}^{z} \\ \Re(N_{ex}) \\ \Im(N_{ex}) \\ \Re(F_{ex}^z) \\ \Im(F_{ex}^z) \end{pmatrix}  , \ 
\mathcal{F}(\mathcal{U}) = \begin{pmatrix}
F_{ee}^z \\ \chi(f_{ee}) N_{ee} \\ F_{xx}^z \\ \chi(f_{xx}) N_{xx} \\ \Re(F_{ex}^z) \\ \Im(F_{ex}^z) \\ \kappa \Re(N_{ex}) + \eta \Re(F_{ex}^z) \\ \kappa \Im(N_{ex}) + \eta \Im(F_{ex}^z) \end{pmatrix} \, .
\end{equation}
In $\mathcal{F}(\mathcal{U})$, the pressure components have been expressed through the closure. We have not included antineutrinos for brevity, as the following results are exactly similar through the replacement $\vrho \to \bvrho$.

First, we focus on the advection problem and verify that the quantum closure leads to physical solutions. At leading order, the flavor-diagonal moments are determined “classically,” such that the Jacobian $\mathbf{J}(\mathcal{U}) \equiv \partial \mathcal{F} / \partial \mathcal{U}$ has a block-triangular structure,
\begin{equation}
\label{eq:jacobian}
    \mathbf{J}(\mathcal{U}) = \left(\begin{array}{c|c|c}
        \mathbf{J}_{\mathrm{class},ee} & \begin{matrix} 0 & 0 \\ 0 & 0 \end{matrix} & \begin{matrix} 0 & 0 & 0 & 0 \\ 0 & 0 & 0 & 0 \end{matrix} \\ \hline
        \begin{matrix} 0 & 0 \\ 0 & 0 \end{matrix} & \mathbf{J}_{\mathrm{class},xx} & \begin{matrix} 0 & 0 & 0 & 0 \\ 0 & 0 & 0 & 0 \end{matrix} \\ \hline
        \begin{matrix} 0 & 0 \\ 0 & 0 \\ * & * \\ * & * \end{matrix} & \begin{matrix} 0 & 0 \\ 0 & 0 \\ * & * \\ * & * \end{matrix} & \begin{matrix} 0 & 0 & 1 & 0 \\
        0 & 0 & 0 & 1 \\
        \kappa & 0 & \eta & 0 \\
        0 & \kappa & 0 & \eta \end{matrix}
\end{array} \right) \, ,
\end{equation}
where
\begin{equation}
    \mathbf{J}_{\text{class},aa} = \begin{pmatrix} 0 & 1 \\ \chi(f_{aa}) - f_{aa} \chi'(f_{aa}) & \chi'(f_{aa}) \end{pmatrix} \, .
\end{equation}
The stars in~\eqref{eq:jacobian} are numbers irrelevant for the calculation of the characteristic speeds of the system~\eqref{eq:QKE}, which are the eigenvalues of $\mathbf{J}(\mathcal{U})$. The first four eigenvalues read
\begin{equation}
    \lambda_{aa}^{(\pm)} = \frac{\chi'(f_{aa}) \pm \sqrt{[\chi'(f_{aa})]^2 - 4 f_{aa} \chi'(f_{aa}) + 4 \chi(f_{aa})}}{2} \, ,
\end{equation}
a well-known result~\cite{Anile_EddingtonFactors_1991,Audit:2002hr}. These speeds are real and satisfy the causality condition $\abs*{\lambda_{aa}^{(\pm)}} \leq 1$, with in particular, in the free-streaming limit, $\lambda_{aa}^{(\pm)} = 1$, which is satisfied as $\chi'(1)=2$ for the Minerbo closure. The last four quantum characteristic speeds are
\begin{equation}
\label{eq:lambda_ex}
    \lambda_{ex}^{(\pm)} = \frac{\eta \pm \sqrt{\eta^2 + 4 \kappa}}{2} \, .
\end{equation}
As we have seen, in the free-streaming limit ($f_{ee},f_{xx} \to 1$) we have $\kappa \to -1$ and $\eta \to 2$. Thus we find $\lambda_{ex}^{(\pm)} \to 1$, satisfying the causality requirement~\cite{Anile_EddingtonFactors_1991,Smit_closure_2000}. More generally, we have numerically checked that these speeds are real and subluminal for all possible values of $\{\vrho_{ee},\vrho_{xx}\}$ which appear in the quantum Minerbo closures (see Appendix~\ref{app:charac_speeds}). In particular, the fact that $\lambda_{ab}^{(\pm)} \in \mathbb{R}$ $\forall (a,b)$ proves that \eqref{eq:QKE} is a \emph{hyperbolic} system, an important basis for future numerical methods.

\begin{figure*}[!ht]
    \centering
    \includegraphics[width=\linewidth]{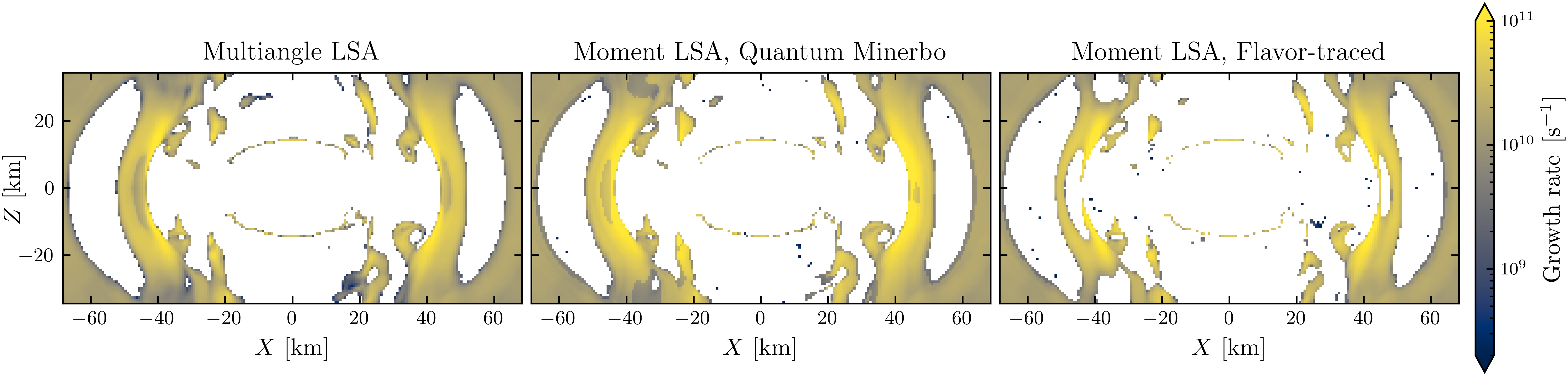}
    \caption{Predicted growth rate of FFIs in a transverse slice of the 5 ms post-merger snapshot for the M1 simulation in~\cite{Foucart:2016rxm}, obtained with multiangle LSA [\emph{left}], two-moment LSA with the quantum closure~\eqref{eq:closure_QMbo} [\emph{center}], and two-moment LSA with the “flavor-traced” closure~\eqref{eq:closure_FT} [\emph{right}].}
    \label{fig:LSA_closures}
\end{figure*}

\section{Linear stability analysis in the “fast” regime}
\label{sec:LSA}

In order to illustrate the improvement gained from the quantum Minerbo closure in actual calculations, we focus on the situation of fast flavor instabilities (FFIs), associated with the existence of an angular crossing between neutrino and antineutrino distributions~\cite{Morinaga:2021vmc,Dasgupta:2021gfs,Fiorillo:2024bzm}. For linear stability analysis (LSA), we are precisely in the small coherence limit assumed in this work: the density matrices are essentially diagonal, with small off-diagonal perturbations seeding the instability. The diagonal elements are taken from a simulation of neutrinos without flavor mixing: here, a general relativistic simulation of the merger of two $1.2 \, M_\odot$ neutron stars, see Ref.~\cite{Foucart:2016rxm}. Given the timescales of the FFI, the source term in Eq.~\eqref{eq:QKE} is reduced to the self-interaction mean-field Hamiltonian.

Generalizing the calculations of Ref.~\cite{Froustey:2023skf}, at each point from the NSM snapshot we look for the fastest growing FFI-unstable mode by undertaking both a multiangle LSA using $n_\mu = 120$ angular bins (see Appendix B in \cite{Froustey:2023skf}), a moment LSA using the “flavor-traced closure” described in \cite{Froustey:2023skf}, and our new quantum Minerbo closure. To accommodate the axisymmetry assumption of this work, at each location of the NSM slice, fluxes are rotated so that the net ELN flux is along the $z$ direction, and we then take $F_{ab}^{x} = F_{ab}^{y} = 0$. In multiangle LSA, we adopt an exponential ansatz for the $2 \times n_\mu$ variables $\vrho_{ex}(\mu_n), \bvrho_{xe}(\mu_n) \propto e^{\i(\Omega t - k z)}$, and we look for the largest value of $\mathrm{Im}(\Omega)$, scanning over wavenumbers $k$. In moment LSA, the same ansatz is adopted for the variables $\{N_{ex},F_{ex}^z,\bN_{xe},\bF_{xe}^z\}$. This requires an expression of $P_{ex}^{zz}$ as a function of $(N_{ex},F_{ex}^z)$, which is given by the closure. Previous dynamical quantum two-moment calculations~\cite{Grohs:2022fyq,Grohs:2023pgq} used an \emph{ad hoc} prescription, designed to “average” the behavior of $\nu_e$ and $\nu_x$ into $\vrho_{ex}$. Namely, for an axisymmetric system, the classical closure $P_{aa}^{zz} = \chi(f_{aa}) N_{aa}$ was used for the flavor on-diagonal moments, while the off-diagonal components followed a “flavor-traced” version of the same classical closure:
\begin{equation}
\label{eq:closure_FT}
    P_{ex}^{zz} = \chi\left(\frac{\lvert F_{ee}^z + F_{xx}^z\rvert}{N_{ee}+N_{xx}}\right) N_{ex} \, .
\end{equation}

The results of our calculations are shown on Fig.~\ref{fig:LSA_closures}. Overall, as evidenced in~\cite{Froustey:2023skf}, the flavor-traced closure shows good performance in identifying the locations and characteristics of FFIs.  However, one can identify several “missing regions,” where the procedure was unable to find the actual instability. On the other hand, the results with the quantum Minerbo closure identify almost perfectly the regions of instability. Generally speaking, moment methods have more difficulty in shallow crossing regions, where the precise choice of closure is of prime importance: we show here the promising features of our new closure prescriptions, with a detailed study of the impact of closure choices in various crossing landscapes left for future work.

\section{Summary \& Prospects}
In this paper we have derived the first quantum generalization of an analytic closure for moment neutrino transport using the maximum entropy principle in the small flavor coherence limit. 

We confirmed that it approaches the correct behavior in the free-streaming and isotropic limits, and that it respects causality, key features needed to formulate well-posed quantum M1 models, which will require further numerical and mathematical work. Beyond the theoretical appeal of such a model, we showed the significantly better capability of this closure in correctly identifying the occurrence of FFIs in a neutron star merger, a requirement for the generalized use of quantum moment methods. We thus showed a close resemblance of the results of moment LSA using the quantum Minerbo closure compared to the multiangle LSA (Fig.~\ref{fig:LSA_closures}), an improvement over alternative, \emph{ad hoc}, closures used previously.
Moreover, a recent study~\cite{Fiorillo:2024qbl} suggests that, in an actual environment, neutrinos may never venture far from stability, in which case the approximation of small flavor coherence that we took in this work is well motivated—although future work will be needed to assess how closures work in situations where the instability is resonant~\cite{Fiorillo:2024bzm,Fiorillo:2024uki}. 

As the first example of a quantum closure based on first principles, this work opens the way to more complex closures (see also~\cite{Kneller:2024buy}), and represents an important step toward the development of robust quantum moment methods for neutrino transport.

\medskip

\begin{acknowledgments}
    
We thank E. Grohs, S. Richers and F. Foucart for many useful discussions and for comments on the manuscript, and in particular F. Foucart for providing the NSM simulation data. J.F. is supported by the Network for Neutrinos, Nuclear Astrophysics and Symmetries (N3AS), through the National Science Foundation Physics Frontier Center award No. PHY-2020275. J.P.K and G.C.M are supported by the United States Department of Energy, Office of Science, Office of Nuclear Physics (Award Nos. DE-FG02-02ER41216 and DE-SC0024388).

\end{acknowledgments}

\appendix

\section{Derivation of the angular distribution~\eqref{eq:distrib_QMbo}}
\label{app:extremization}

We give here additional details on the maximization of the functional~\eqref{eq:functional_quant}, leading to the angular distribution $\vrho_{ex}(\mu)$ given in Eq.~\eqref{eq:distrib_QMbo}. First, we note that the Frobenius inner products that set the constraints read explicitly

\begin{widetext}
\begin{multline}
    \left \langle \alpha, \int_{-1}^{1}{\mathrm{d}\mu \, \vrho} - \frac{N}{2 \pi}\right\rangle = \alpha_{ee} \left(\int_{-1}^{1}{\mathrm{d}\mu \, \vrho_{ee}} - \frac{N_{ee}}{2 \pi} \right) + \alpha_{xx} \left(\int_{-1}^{1}{\mathrm{d}\mu \, \vrho_{xx}} - \frac{N_{xx}}{2 \pi} \right) \\
    + 2 \, \mathrm{Re}(\alpha_{ex}) \left(\int_{-1}^{1}{\mathrm{d}\mu \, \mathrm{Re}(\vrho_{ex})} - \frac{\mathrm{Re}(N_{ex})}{2 \pi} \right) + 2 \, \mathrm{Im}(\alpha_{ex}) \left(\int_{-1}^{1}{\mathrm{d}\mu \, \mathrm{Im}(\vrho_{ex})} - \frac{\mathrm{Im}(N_{ex})}{2 \pi} \right) \, ,
\end{multline}
\end{widetext}

with the same expression for the flux constraint with $\alpha \to Z$, $N \to F^z$ and $\mathrm{d}\mu \to \mu \, \mathrm{d}\mu$. We recall that the entropy term reads [see Eq.~\eqref{eq:Tr_rhologrho}]
\begin{multline}
    \mathrm{Tr}[\vrho \ln(\vrho)] = \vrho_{ee} \ln(\vrho_{ee}) + \vrho_{xx} \ln(\vrho_{xx}) \\
    + \frac{\ln(\vrho_{ee})-\ln(\vrho_{xx})}{\vrho_{ee} - \vrho_{xx}} \left[\mathrm{Re}(\vrho_{ex})^2 + \mathrm{Im}(\vrho_{ex})^2\right] \, .
\end{multline}
Therefore, maximizing the functional $\mathsf{S}[\vrho]$ with respect to $\mathrm{Re}(\vrho_{ex})$ gives
\begin{equation}
    0 = \frac{\ln(\vrho_{ee})-\ln(\vrho_{xx})}{\vrho_{ee} - \vrho_{xx}} 2 \, \mathrm{Re}(\vrho_{ex}) - 2 \,  \mathrm{Re}(\alpha_{ex}) - 2 \, \mu \, \mathrm{Re}(Z_{ex}) \, ,
\end{equation}
hence
\begin{equation}
    \mathrm{Re}(\vrho_{ex}) = \frac{\vrho_{ee} - \vrho_{xx}}{\ln(\vrho_{ee}) - \ln(\vrho_{xx})} \left[ \mathrm{Re}(\alpha_{ex}) + \mu \mathrm{Re}(Z_{ex})\right] \, .
\end{equation}
Repeating the same procedure for $\mathrm{Im}(\vrho_{ex})$ and summing $\mathrm{Re}(\vrho_{ex}) + \i\, \mathrm{Im}(\vrho_{ex}) = \vrho_{ex}$ leads to the expression~\eqref{eq:distrib_QMbo}.

\section{Quantum Fermi-Dirac maximum entropy closure}
\label{app:FermiDirac}

We derive here the fermionic version of our quantum maximum entropy closure, first derived in the classical case in~\cite{Cernohorsky_1989}.

Taking into account the fermionic nature of neutrinos, the functional~\eqref{eq:functional_quant} becomes
\begin{multline}
\label{eq:functional_quant_FD}
    \mathsf{S}[\vrho] = \int_{-1}^{1}{\dd \mu \, \Tr\left[\vrho \ln(\vrho) + (\Id-\vrho)\ln(\Id-\vrho)\right]} \\ - \left \langle \alpha, \int_{-1}^{1}{\dd \mu  \vrho} - \frac{N}{2 \pi \vrho_0} \right\rangle - \left \langle Z, \int_{-1}^{1}{\dd \mu \mu \vrho} - \frac{F^z}{2 \pi \vrho_0}\right\rangle \, ,
\end{multline}
where $\vrho_0$ is a normalization factor, needed to ensure that $\vrho_{aa}$ is a distribution function with $0 \leq \vrho_{aa} \leq 1$. 

The additional term compared to the Maxwell-Boltzmann limit~\eqref{eq:functional_quant} gives a contribution identical to \eqref{eq:Tr_rhologrho} with $\vrho_{aa} \to 1 - \vrho_{aa}$, that is,
\begin{widetext}
\begin{equation}
\label{eq:Tr_1mrholog1mrho}
    \Tr\left[(\Id - \vrho)\ln(\Id -\vrho)\right] = (1-\vrho_{ee})\ln(1- \vrho_{ee}) + (1-\vrho_{xx})\ln(1-\vrho_{xx}) - \frac{\ln(1-\vrho_{ee}) - \ln(1-\vrho_{xx})}{\vrho_{ee}-\vrho_{xx}} \abs*{\vrho_{ex}}^2 + \cdots
\end{equation}
For completeness, we quote the same formulas, but written for any number of flavors:\footnote{The prefactor of $1/2$ in front of the second sum disappears in~\eqref{eq:Tr_1mrholog1mrho} since $\vrho_{xe}=\vrho_{ex}^*$.}
\begin{align}
    \Tr\left[\vrho \ln(\vrho)\right] &= \sum_{a}{\vrho_{aa} \ln(\vrho_{aa})} + \frac12 \sum_{a \neq b}{\frac{\ln(\vrho_{aa})-\ln(\vrho_{bb})}{\vrho_{aa}-\vrho_{bb}}\abs*{\vrho_{ab}}^2} + \cdots \, , \\
    \Tr\left[(\Id-\vrho) \ln(\Id-\vrho)\right] &= \sum_{a}{(1-\vrho_{aa}) \ln(1-\vrho_{aa})} - \frac12 \sum_{a \neq b}{\frac{\ln(1-\vrho_{aa})-\ln(1-\vrho_{bb})}{\vrho_{aa}-\vrho_{bb}}\abs*{\vrho_{ab}}^2} + \cdots \, .
\end{align}
\end{widetext}

\subsubsection*{Flavor-diagonal elements}

Maximizing~\eqref{eq:functional_quant_FD} over $\vrho_{aa}$ yields, at leading order,
\begin{equation}
    \ln(\vrho_{aa}) - \ln(1-\vrho_{aa}) - \alpha_{aa} \vrho_{aa} - Z_{aa} \, \mu \, \vrho_{aa} = 0 \, ,
\end{equation}
such that
\begin{equation}
    \frac{\vrho_{aa}}{1-\vrho_{aa}} = e^{\alpha_{aa}}e^{Z_{aa}\mu} \, , 
\end{equation}
or equivalently
\begin{equation}
\vrho_{aa} = \frac{1}{e^{-\alpha_{aa}}e^{-Z_{aa}\mu} + 1} \, .
\end{equation}
The constraints read
\begin{equation}
        2 \pi \int_{-1}^{1}{\dd \mu \begin{bmatrix} 1 \\ \mu \end{bmatrix} \frac{\vrho_0}{e^{-\alpha_{aa}}e^{-Z_{aa}\mu}+1}} = \begin{bmatrix} N_{aa} \\ F_{aa}^z \end{bmatrix} \, .
\end{equation}
They were historically numerically inverted or considered in a Padé approximation, before Cernohorsky and Bludman~\cite{Cernohorsky_closure_1994} found an approximate closed-form for the Eddington factor $P_{aa}^{zz}/N_{aa}$ (see also~\cite{Smit_closure_2000,Murchikova:2017zsy}).

\subsubsection*{Flavor off-diagonal elements} 

We now maximize~\eqref{eq:functional_quant_FD} with respect to $\Re(\vrho_{ex})$ and $\Im(\vrho_{ex})$. The result is straightforward given the Maxwell-Boltzmann derivation in Appendix~\ref{app:extremization}, as the term in $\mathsf{S}[\vrho]$ involving $\vrho_{ex}$ that changes because of Fermi-Dirac statistics reads:
\begin{equation}
    \mathsf{S}[\vrho] \supset \frac{\ln\left(\dfrac{\vrho_{ee}}{1-\vrho_{ee}}\right) - \ln\left(\dfrac{\vrho_{xx}}{1-\vrho_{xx}}\right)}{\vrho_{ee}-\vrho_{xx}} \abs{\vrho_{ex}}^2 \, ,
\end{equation}
such that Eq.~\eqref{eq:distrib_QMbo} becomes
\begin{equation}
    \vrho_{ex}(\mu) = \frac{\vrho_{ee}-\vrho_{xx}}{\ln\left(\dfrac{\vrho_{ee}}{1-\vrho_{ee}}\right) - \ln\left(\dfrac{\vrho_{xx}}{1-\vrho_{xx}}\right)}\left(\alpha_{ex} + \mu Z_{ex}\right) \, .
\end{equation}
Therefore, Eqs.~\eqref{eq:alpha_Z}--\eqref{eq:kappa_eta} keep the same form, where the $I_{(n)}$ integrals are modified via $\vrho_{aa}\to \vrho_{aa}/(1-\vrho_{aa})$ inside the logarithms. Notably, the angular dependence of the denominator of $I_{(n)}$ remains linear in $\mu$, just like in the Maxwell-Boltzmann case. 

\section{Closure parameters in the free-streaming regime}
\label{app:free_streaming}

In this Appendix, we provide a derivation of the limits of the closure coefficients $\kappa$ and $\eta$ when the flux factors for $\nu_e$ and $\nu_x$ approach $1$.

When $f_{ee} \to 1$, $Z_{ee} \to + \infty$ according to Eq.~\eqref{eq:Z_ME}, such that $\vrho_{ee}(\mu) \sim (N_{ee}/2 \pi) Z_{ee} e^{Z_{ee}(\mu-1)}$. This is a function that, for $Z_{ee} \to + \infty$, is nonzero only in a small neighborhood of $\mu = 1$. The same arguments apply to $\vrho_{xx}$. Recalling that
\begin{equation}
\begin{aligned}
    I_{(n)} &= 2 \pi \int_{-1}^{1}{\dd \mu \, \mu^n \frac{\vrho_{ee} - \vrho_{xx}}{\ln(\vrho_{ee})-\ln(\vrho_{xx})}} \\
    &\equiv 2 \pi \int_{-1}^{1}{\dd \mu \, \mu^n \, h(\mu)} \, ,
\end{aligned}
\end{equation}
we distinguish different cases.

First, if $\vrho_{ee}(\mu)=\vrho_{xx}(\mu)$, since the limit of $x \mapsto (x-1)/\ln(x)$ for $x \to 1$ is $1$, then $h(\mu) = \vrho_{ee}(\mu)$. 

Otherwise, we have $N_{ee} \neq N_{xx}$ and/or $Z_{ee} \neq Z_{xx}$, such that the denominator of $h(\mu)$ reads, in the $f_{aa}\to 1$ limit, $\ln(N_{ee}/N_{xx}) + (Z_{ee}-Z_{xx})\mu$, which is a slowly varying function of $\mu$ compared to the exponentials in the numerator. Since the support of $I_{(n)}$ is very localized at $\mu \lesssim 1$, we can take the denominator as constant. Assuming, without loss of generality, that $Z_{xx} > Z_{ee} \gg 1$, we have
\begin{equation}
    \vrho_{ee} - \vrho_{xx} = \vrho_{ee} \left(1 - \frac{\vrho_{xx}}{\vrho_{ee}} \right) \sim \vrho_{ee} \, ,
\end{equation}
as $\vrho_{xx}/\vrho_{ee} \sim (N_{ee}Z_{ee}/N_{xx}Z_{xx})e^{-(Z_{xx}-Z_{ee})(1-\mu)} \ll 1$, $\forall \mu < 1$.

Consequently, in any case, $I_{(n)}$ is given, in the free-streaming limit, by $I_{(n)} = C \times J_{(n)}(1/Z_{ee})$, with $C$ a constant independent of $n$, and
\begin{equation}
    J_{(n)}(\veps) \equiv \int_{-1}^{1}{\dd \mu \, \mu^n \frac{1/\veps}{2 \sinh(1/\veps)}e^{\mu/\veps}} \, .
\end{equation}
In particular, this means that the limits of $\kappa$ and $\eta$, defined by Eq.~\eqref{eq:kappa_eta}, are given by
\begin{equation}
\label{eq:limit_kappa_eta}
\begin{aligned}
 \lim_{f_{ee},f_{xx} \to 1} \kappa &= \lim_{\veps \to 0} \frac{J_{(2)}^2 - J_{(1)}J_{(3)}}{J_{(0)}J_{(2)}- J_{(1)}^2} \, , \\
\lim_{f_{ee},f_{xx} \to 1} \eta &= \lim_{\veps \to 0} \frac{J_{(0)}J_{(3)} - J_{(1)}J_{(2)}}{J_{(0)}J_{(2)}- J_{(1)}^2} \, .
\end{aligned}
\end{equation}
We have, for $Z_{ee} \to \infty$ and thus $\veps \to 0$,
\begin{equation}
    \begin{aligned}
        J_{(0)} &= 1 \, , \quad & \quad J_{(2)} &\sim 1 - 2 \veps + 2 \veps^2 \, , \\
        J_{(1)} &\sim 1 - \veps  \, , \quad & \quad J_{(3)} &\sim 1 - 3 \veps + 6 \veps^2 - 6 \veps^3 \, .
    \end{aligned}
\end{equation}
As a consequence,
\begin{equation}
\label{eq:limit_distribs}
    \begin{aligned}
        J_{(0)}J_{(2)} - J_{(1)}^2 &\sim \veps^2 \, , \\
        J_{(2)}^2 - J_{(1)}J_{(3)} &\sim - \veps^2 \, , \\
        J_{(0)}J_{(3)} - J_{(1)}J_{(2)} &\sim 2 \veps^2 \, .
    \end{aligned}
\end{equation}
Inserting those asymptotic equivalents in Eq.~\eqref{eq:limit_kappa_eta}, we proved that $\kappa \to -1$ and $\eta \to 2$. If the flux factors tend to $-1$, the support of $J_{(n)}$ is around $-1$ and there is an extra minus sign in front of all the odd $J_{(n)}$, such that $\eta \to -2$.

\begin{figure*}[ht]
    \centering
    \includegraphics[width=0.96\linewidth]{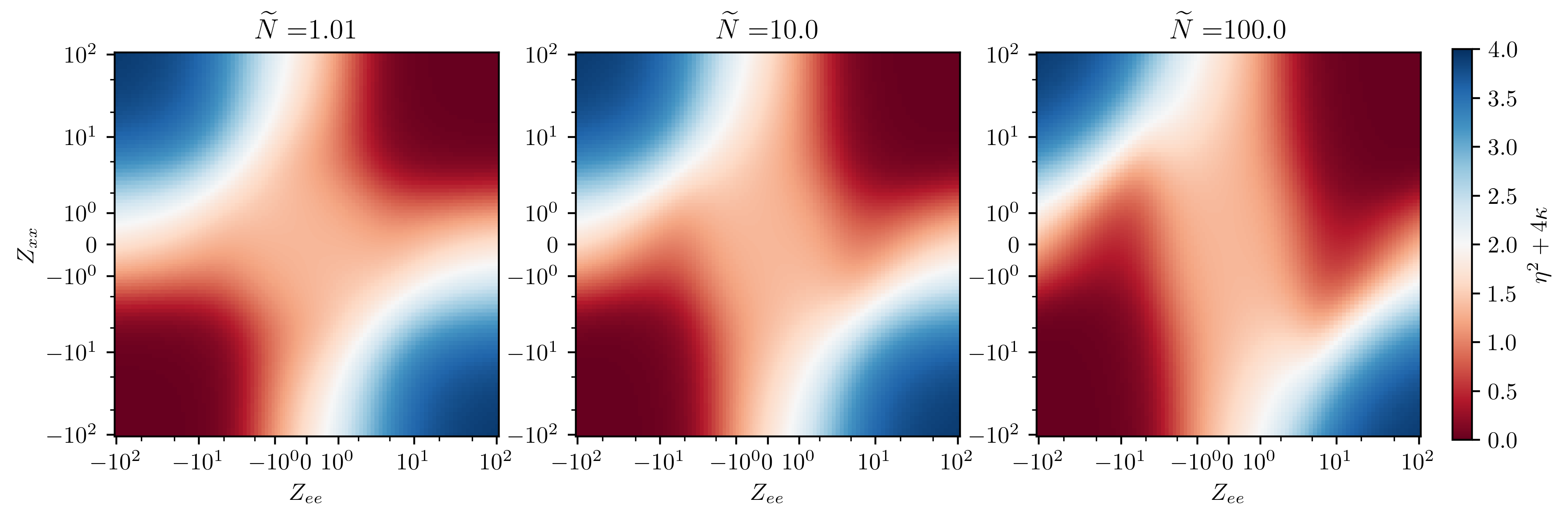}
    \caption{Combination $\eta^2 + 4 \kappa$ across the range of values of $\{Z_{ee},Z_{xx}\}$, for three relative number densities $\widetilde{N} \in \{1.01,10,100\}$.}
    \label{fig:check_lambda_real}
\end{figure*}

\section{Numerical study of $\lambda_{ex}^{(\pm)}$}
\label{app:charac_speeds}

In order to verify that the additional characteristic speeds~\eqref{eq:lambda_ex} are real and satisfy the causality requirements $\abs*{\lambda_{ex}^{(\pm)}} \leq 1$, we perform a numerical scan over the parameter space that can be spanned by the Minerbo distributions entering Eq.~\eqref{eq:kappa_eta}.

First, assuming without loss of generality that $N_{ee}\geq N_{xx}$, we rewrite the $I_{(n)}$ integrals~\eqref{eq:In} as
\begin{widetext}
\begin{equation}
    I_{(n)} = \frac{N_{xx}}{2} \int_{-1}^{1}{\dd{\mu} \, \mu^n \frac{\widetilde{N} \dfrac{Z_{ee}}{\sinh(Z_{ee})} e^{Z_{ee} \mu} - \dfrac{Z_{xx}}{\sinh(Z_{xx})}e^{Z_{xx} \mu}}{\ln\left[\widetilde{N} \dfrac{Z_{ee} \sinh(Z_{xx})}{Z_{xx} \sinh(Z_{ee})}\right] + (Z_{ee}-Z_{xx})\mu}} \, ,
\end{equation}
\end{widetext}
with $\widetilde{N} = N_{ee}/N_{xx}$. Since ratios of $I_{(n)}$ appear in the functions $\kappa$ and $\eta$, these quantities (and the related characteristic speeds $\lambda_{ex}^{(\pm)}$) are functions of three parameters: $\widetilde{N} \in [1,+\infty)$, $Z_{ee} \in \mathbb{R}$, and $Z_{xx} \in \mathbb{R}$ (note that a negative $Z_{aa}$ corresponds to $F_{aa}^{z} \leq 0$).

\begin{figure*}[ht]
    \centering
    \includegraphics[width=0.98\linewidth]{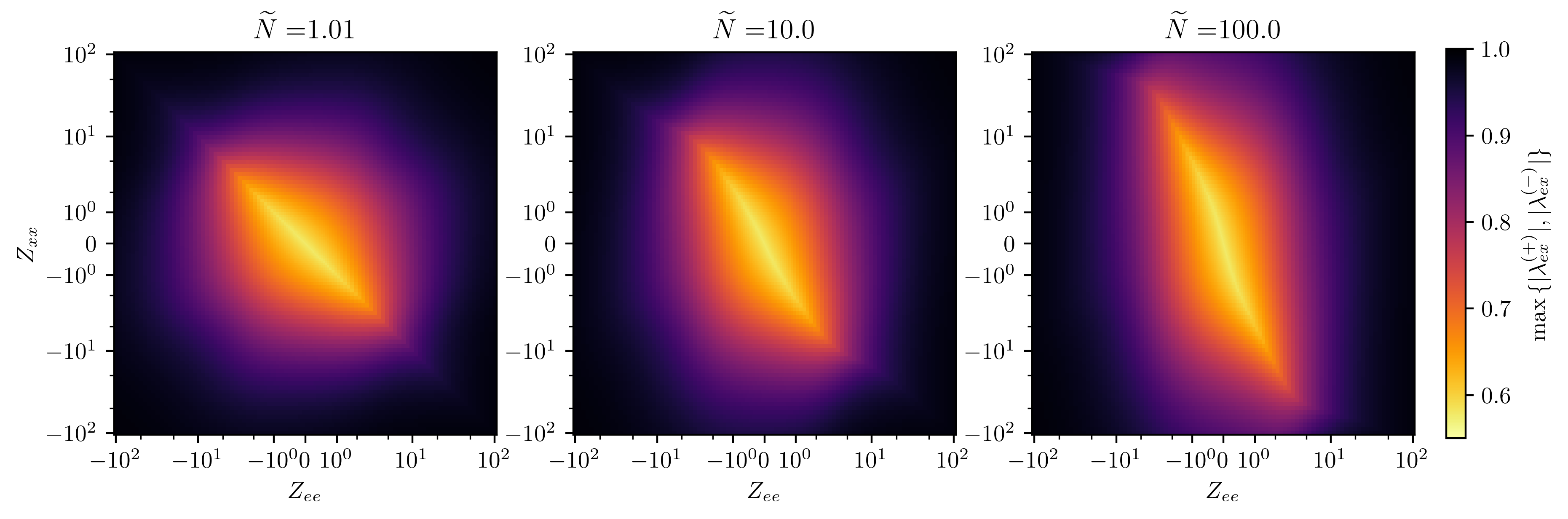}
    \caption{Maximum $\lambda_{ex}$ characteristic speed for the same range of parameters as Fig.~\ref{fig:check_lambda_real}.}
    \label{fig:scan_params_lambda}
\end{figure*}

In Fig.~\ref{fig:check_lambda_real}, we scan over the parameter ranges and display the quantity $\eta^2 + 4 \kappa$, which must be positive so that $\lambda_{ex}^{(\pm)} \in \mathbb{R}$ [see Eq.~\eqref{eq:lambda_ex}]. In Fig.~\ref{fig:scan_params_lambda} we plot the maximum quantum characteristic speed, which must be below $1$. The minimum value of $\eta^2 + 4 \kappa$ is obtained when $Z_{ee}$ and $Z_{xx}$ both tend toward $\pm \infty$ together, which corresponds to two beams of neutrinos in the same direction. We have proven in Appendix~\ref{app:free_streaming} [see after Eq.~\eqref{eq:limit_distribs}] that in this situation $\kappa \to -1$ and $\eta \to 2$, which proves that $\eta^2 + 4 \kappa \to 0$. As a consequence, the eigenvalues of the Jacobian matrix~\eqref{eq:jacobian} are all real, such that the moment system is hyperbolic~\cite{Pons:2000br,Tadmor_2016}.

When $\abs*{Z_{aa}} \gg 1$, meaning that $\nu_a$ is free-streaming, we see on Fig.~\ref{fig:scan_params_lambda} that the maximum characteristic speed goes to $1$, as expected by causality requirements~\cite{Anile_EddingtonFactors_1991}. When both $\abs*{Z_{ee}}, \abs*{Z_{xx}} \ll 1$, this is the isotropic limit for which $\eta \to 0$ and $\kappa \to 1/3$, so that $\abs*{\lambda_{ex}^{(\pm)}} \to 1/\sqrt{3} \simeq 0.577$.

We thus explicitly show that characteristic speeds are everywhere physical, an important requirement for a closure to be used in large-scale hydrodynamic simulations.

\begin{figure}[!ht]
    \centering
    \includegraphics[width=\linewidth]{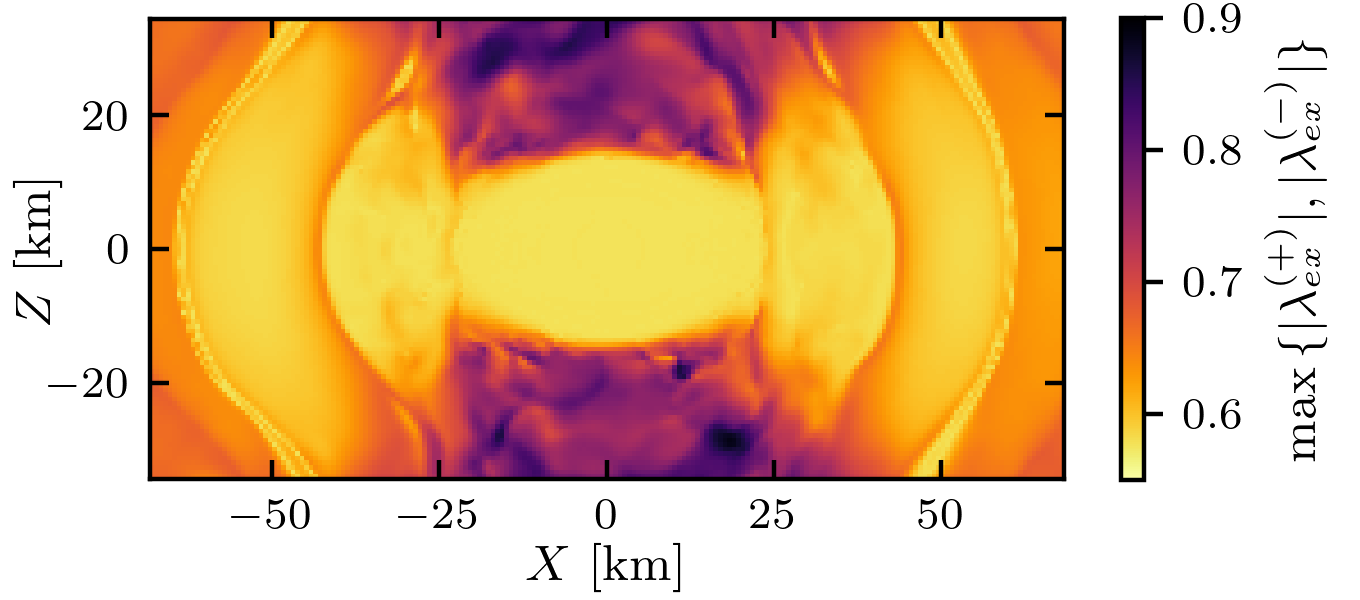}
    \caption{Largest characteristic speed $\lvert\lambda_{ex}^{(\pm)}\rvert$ in a transverse slice of the 5 ms postmerger snapshot from the NSM simulation~\cite{Foucart:2016rxm}.}
    \label{fig:lambda_NSM}
\end{figure}

As an additional illustration, we show on Fig.~\ref{fig:lambda_NSM} the largest quantum characteristic speed at each point in the transverse slice of the 5 ms postmerger snapshot from \cite{Foucart:2016rxm}, studied in Sec.~\ref{sec:LSA} (see Fig.~\ref{fig:LSA_closures}). As proved above, the characteristic speeds are always below unity, approaching the isotropic limit in the center of the slice ($\lambda_{ex}^{(\pm)} \to \pm 1/\sqrt{3} \simeq 0.577$). The characteristic speeds do not get significantly close to $1$ in this snapshot, as the simulation box is rather close to the remnant.

\bibliography{references}

\end{document}